\documentclass[a4paper,12pt]{article}

\newcommand{\sect}[1]{\setcounter{equation}{0}\section{#1}}

\begin{document}
\topmargin 0pt \oddsidemargin 0mm

\renewcommand{\thefootnote}{\fnsymbol{footnote}}
\begin{titlepage}
\begin{flushright}
\end{flushright}

\vspace{1mm}
\begin{center}
{\Large \bf Holographic Confinement/Deconfinement Phase Transitions
of AdS/QCD in Curved Spaces}

 \vspace{10mm}
{\large Rong-Gen Cai\footnote{Email address: cairg@itp.ac.cn} and
Jonathan P. Shock\footnote{Email address:
 jps@itp.ac.cn}}\\
\vspace{5mm}
{ \em Institute of Theoretical Physics, Chinese Academy of Sciences, \\
   P.O. Box 2735, Beijing 100080, China}\\

\end{center}
\vspace{5mm} \centerline{{\bf{Abstract}}}
 \vspace{5mm}
 Recently Herzog has shown that deconfinement of AdS/QCD can
 be realized, in the hard-wall model where the small radius region
is removed in the asymptotically AdS space, via a first order
Hawking-Page phase transition between a low temperature phase given
by a pure AdS geometry and a high temperature phase given by the AdS
black hole in Poincare coordinates. In this paper we first extend
Herzog's work to the hard wall AdS/QCD model in curved spaces by
studying the thermodynamics of AdS black holes with spherical or
negative constant curvature horizon, dual to a non-supersymmetric
Yang-Mills theory on a sphere or hyperboloid respectively. For the
spherical horizon case, we find that the temperature of the phase
transition increases by introducing an infrared cutoff, compared to
the case without the cutoff; For the hyperbolic horizon case, there
is a gap for the infrared cutoff, below which the Hawking-Page phase
transition does not occur. We also discuss charged AdS black holes
in the grand canonical ensemble, corresponding to a Yang-Mills
theory at finite chemical potential, and find that there is always a
gap for the infrared cutoff due to the existence of a minimal
horizon for the charged AdS black holes with any horizon topology.

\end{titlepage}

\newpage
\renewcommand{\thefootnote}{\arabic{footnote}}
\setcounter{footnote}{0} \setcounter{page}{2}
\sect{Introduction}

The remarkable AdS/CFT  correspondence~\cite{AdS} conjectures that
string/M theory in an anti-de Sitter space (AdS) times a compact
manifold is dual to a large $N$ strongly coupling conformal field
theory (CFT) residing on the boundary of the AdS space. A special
example of the AdS/CFT correspondence is that type IIB string theory
in $AdS_5\times S^5$ is dual to a four dimensional ${\cal N}=4$
supersymmetric Yang-Mills theory on the boundary of $AdS_5$. At
finite temperature, in the spirit of the AdS/CFT correspondence,
Witten~\cite{Witten} has argued that the thermodynamics of black
holes in AdS space can be identified with that of the dual strongly
coupling field theory in the high temperature limit. Therefore one
can discuss the thermodynamics and phase structure of strongly
coupling field theories by studying the thermodynamics of various
kinds of black holes in AdS space. Indeed, it is well-known that
there exists a phase transition between the Schwarzschild-AdS black
hole and thermal AdS space, the so-called Hawking-Page phase
transition~\cite{HP}: the black hole phase dominates the partition
function in the high temperature limit, while the thermal AdS space
dominates in the low temperature limit. This phase transition is of
first order, and is interpreted as the confinement/deconfinement
phase transition in the dual field theory~\cite{Witten}. Note that
the dual conformal field theory to the Schwarzschild-AdS black hole
configuration resides on a sphere. For example, for the five
dimensional Schwarzschild-AdS black holes, the dual field theory is
${\cal N}=4$ supersymmetric Yang-Mills theory at finite
temperature~\cite{Witten}. That is, by discussing the thermodynamics
of AdS black holes with a spherical horizon, one can study
thermodynamics of the dual non-supersymmetric Yang-Mills theory on a
sphere at finite temperature.

It is interesting to note that in AdS space the black hole horizon
is not necessarily a sphere~\cite{Topo}, the black hole horizon can
also be a Ricci flat surface \cite{Flat} or a negative constant
curvature surface~\cite{Negative}. Thus using those AdS black hole
solutions, one can study dual strongly coupled field theories
residing on Ricci flat or hyperbolic spaces. These so-called
topological black holes have been investigated in higher dimensions
\cite{Birm,high,GB,Love} and in dilaton gravity~\cite{CJS,CZ}. It
was found that the Hawking-Page phase transition, which happens for
spherical AdS black holes, does not occur for Ricci flat and
negative curvature AdS black holes, the latter two being not only
locally stable (heat capacity is always positive), but also globally
stable (see for example, \cite{Birm}). Here it is worth mentioning
that if some directions of the horizon surface are compact for Ricci
flat AdS black holes, similar to the case of spherical AdS black
holes, the Hawking-Page phase transition can
occur~\cite{SSW,Page,Myers,CKW,BD} due to the existence of so-called
AdS soliton~\cite{HM}. The absence of a Hawking-Page phase
transition for Ricci flat AdS black holes without compact directions
is consistent with the result that the AdS space and the AdS black
holes in Poincare coordinates without compact directions are both in
the deconfinement phases~\cite{Witten}. This can be confirmed by
calculating quark-anti-quark potential through
Wilson-loop~\cite{Wilson}. In addition, the black hole entropy is
proportional to $N^2$, where $N$ is the rank of gauge group $SU(N)$
for the dual gauge field.

Using the AdS/CFT correspondence, it is expected that we can obtain
some more qualitative understanding of QCD and the nature of
confinement. The authors of papers~\cite{KS,MN,PS} are able to
realize confinement of related supersymmetric field theories by
finding dual gravitational configurations where the geometry in the
infrared at small radius is capped off in a smooth way. The author
of \cite{PS2} proposed a simpler model, $AdS_5$ in Poincare
coordinates without compact directions where the small radius region
is removed, to realize confinement. The field theory dual to this is
a non-supersymmetric Yang-Mills theory in four dimensions. Although
such a model is somewhat rough, the results obtained in
\cite{EKSS,DP} show that one can get some realistic,
semiquantitative descriptions of low energy QCD, by using this hard
wall model.

Recently, Herzog~\cite{Herz} (see also \cite{BBBZ}) has shown that
in this simple hard wall model of AdS/QCD, the
confinement/deconfinement phase transition occurs via the first
order Hawking-Page phase transition between the low temperature
thermal AdS space and high temperature AdS black hole in Poincare
coordinates. Note the facts that removing the small radius region of
$AdS_5$ is dual to introducing an infrared (IR) cutoff in the dual
field theory and that the Hawking-Page phase transition does not
occur for the negative constant curvature AdS black holes without a
cutoff. It is quite interesting to revisit the thermodynamics and
the Hawking-Page phase transitions for spherical AdS black holes and
negative constant curvature AdS black holes by removing the small
radius regions. That is, we are interested in the
confinement/deconfinement phase transition for the hard wall QCD
model in curved spaces. Recall that charged AdS black hole
configurations are dual to supersymmetric field theories with
so-called R-charges~\cite{CEJM,CG,Cai}. Introducing R-charges to the
AdS black holes is equivalent to introducing a chemical potential in
the dual field theory. In this paper, therefore, we will also
discuss the case of charged AdS black holes.

The paper is organized as follows. In the next section we will
revisit the Hawking-Page phase transition for spherical AdS black
holes with an IR cutoff. The case for the negative constant
curvature black holes will be discussed in Sec.~3. In. Sec.~4 we
will investigate the phase transition for charged AdS black holes
with an IR cutoff. Sec.~5 will be devoted to conclusions and
discussions.


\sect{Hawking-Page Phase Transition for Spherical Black Holes with
an IR Cutoff}

Let us start with the action of five-dimensional general relativity
with a negative cosmological constant
\begin{equation}
\label{2eq1}
 {\cal S}= \frac{1}{2\kappa^2} \int d^5x \sqrt{-g} \left(
 R+\frac{12}{l^2}\right),
 \end{equation}
 where $k^2= 8\pi G$ and $l$ is the radius of the five-dimensional
 AdS space.
 The action (\ref{2eq1}) admits the AdS spherical black hole
 solution with the metric
 \begin{equation}
 \label{2eq2}
  ds^2= -V(r) dt^2 + V(r)^{-1} dr^2 + r^2 d\Omega_3^2,
  \end{equation}
  where
  \begin{equation}
  \label{2eq3}
  V(r) = 1+ \frac{r^2}{l^2} -\frac{r^2_s}{r^2},
  \end{equation}
  $d\Omega_3^2$ is the line element of a three-dimensional sphere
  with unit radius and $r_s^2$ is an integration constant,
  the Schwarzschild mass parameter. The black hole horizon is determined
  by $V(r_+)=0$, which gives $r_+^2 = l^2
  (-1+\sqrt{1+4r_s^2/l^2})/2$.  The black hole has an inverse temperature $1/T$
  \begin{equation}
 \beta = 1/T=\frac{2\pi r_+}{1+ 2r_+^2/l^2}.
 \end{equation}
  In the global coordinate
  (\ref{2eq2}), the AdS boundary is located at $r\rightarrow
  \infty$. In Poincare coordinates the radial direction is dual to the energy scale of
  the dual field theory. If one introduces a regulating UV cutoff at $r=R \gg r_+$, then
  the range from $r=R$ to $ \infty$ should be removed from the bulk
geometry. On the other hand, one further introduces an IR
  cutoff at $r=r_0$ in the dual field theory (The IR cutoff is
  equivalent to a mass gap in the dual field theory), the range from $r=0$ to $r_0$
  also should be removed from the bulk geometry~\cite{PS2}. This corresponds to a hard wall
  truncation of the space, independent of the radius of the horizon of the black
  hole. Namely, we are
  considering the bulk geometry from $r=r_0$ to $r=R$. The UV cutoff
  is necessary in order to regularize the action. At the end of
  calculation, we will remove the UV cutoff by letting
  $R\rightarrow \infty$.

 To see the phase structure of QCD in the hard wall AdS/QCD
 model, let us calculate the Euclidean action of the AdS
 black hole, by choosing the AdS vacuum solution (\ref{2eq2}) with
\begin{equation}
\label{2eq5}
 V_b=1+\frac{r^2}{l^2}
 \end{equation}
 as the reference background. As a result, the Euclidean action of
 the reference background is
 \begin{equation}
 \label{2eq6}
 {\cal S}_b= \frac{4\Omega_3}{\kappa^2 l^2} \beta_b \int^R _{r_0}
 r^3 dr,
 \end{equation}
 while the Euclidean action of the black hole is
 \begin{equation}
 \label{2eq7}
 {\cal S}_{bh} = \frac{4\Omega_3}{\kappa^2l^2}\beta
 \int^R_{r_{\rm max}}r^3dr,
 \end{equation}
 where $\Omega_3$ is the volume of the unit three-sphere and
 $r_{\rm max}= {\rm max}(r_0,r_+)$ is the IR cutoff in the high temperature phase.
 If $r_0<r_+$, one has $r_{\rm max}=r_+$ while $r_{\rm max}=r_0$ as
 $r_0>r_+$. In addition, let us mention here that usually the
 Hilbert-Einstein action of general relativity should be supplemented
 by the Gibbons-Hawking surface term in order to have a well-defined
 variational principle. However, for the case of the asymptotically AdS spaces, such
 terms have no contribution to the difference of two Euclidean
 actions between the configuration under consideration and the
 reference background~\cite{Witten}.

 In order that the black hole solution (\ref{2eq2}) can be embedded
 into the background consistently, at the boundary $r=R$ the period
 $\beta_b$ of the Euclidean time for the reference background has to
 obey the relation
 \begin{equation}
 \label{2eq8}
 \beta_b \sqrt{V_b(R)}=\beta \sqrt{V(R)}.
 \end{equation}
 Using (\ref{2eq8}), calculating the difference between (\ref{2eq7})
 and (\ref{2eq6}), and taking the limit $ R\rightarrow \infty $, we
 obtain
 \begin{equation}
 \label{2eq9}
 {\cal I} = -\frac{\Omega_3 \beta }{2\kappa^2 l^2} \left ( 2r^4_{\rm
 max }-r_+^4 -2r_0^4 -r_+^2l^2 \right ).
 \end{equation}
Without the IR cutoff, namely $r_0=0$, the action (\ref{2eq9})
reduces to
\begin{equation}
\label{2eq10}
 {\cal I} =-\frac{\Omega_3 \beta }{2\kappa^2 l^2}r_+^2  ( r_+^2 -l^2
 ).
\end{equation}
Clearly when $r_+=l$, the Euclidean action alters its sign, a
first-order phase transition happens. This is just the well-known
Hawking-Page phase transition.  The phase transition temperature is
\begin{equation}
\label{2eq11}
 T_{\rm HP}=\frac{3}{2\pi l}.
\end{equation}
When $r_+>l$, or $T>T_{\rm HP}$, the black hole phase is dominant
in the partition function, while the thermal AdS space is dominant
as $T<T_c$ in the low temperature phase. With the action
(\ref{2eq10}), we obtain the mass of the AdS black hole, via
$E=\partial {\cal I}/\partial \beta$,
\begin{equation}
\label{2eq12}
 E= \frac{3\Omega_3r_+^2}{2\kappa^2}\left(
1+\frac{r_+^2}{l^2}\right)=M,
\end{equation}
and the entropy of the black hole, via $S=\beta E-{\cal I}$,
  \begin{equation}
  \label{2eq13}
S = \frac{\Omega_3r_+^3}{4G},
\end{equation}
satisfying the area formula of black hole entropy.

Now we consider the case with an IR cutoff, $r_0$. There are two
quite different cases.

  (i)  $r_0 \ge r_+$. In this case, one has $r_{\rm max}=r_0$. And the
  action (\ref{2eq9}) turns out to be
  \begin{equation}
  {\cal I}= \frac{\Omega_3 \beta }{2\kappa^2 l^2} \left (
r_+^4 +r_+^2l^2 \right ).
 \end{equation}
 Clearly in this case, the action is always positive and therefore
 no phase transition will occur. The thermal AdS space is globally stable and
 is dominant in the partition function of the dual field theory.

 (ii) $r_0<r_+$. In this case one has $r_{\rm max}=r_+$. And the
 action (\ref{2eq9}) becomes
 \begin{equation}
 \label{2eq15}
 {\cal I} = -\frac{\Omega_3 \beta }{2\kappa^2 l^2} \left (
r_+^4 -2r_0^4 -r_+^2l^2 \right ).
\end{equation}
Obviously, the action changes its sign at
\begin{equation}
r_+^2 = \frac{l^2}{2}\left( 1+\sqrt{1+8r_0^4/l^4} \right).
 \end{equation}
 Therefore in this case, the phase transition can occur and the
 critical temperature is
 \begin{equation}
 T_c = \frac{2+\sqrt{1+8r_0^4/l^4}}{\sqrt{2}\pi l
 \sqrt{1+\sqrt{1+8r_0^4/l^4}}} >T_{\rm HP}.
 \end{equation}
This critical temperature is higher than the one for the case
without the IR cutoff. The thermal energy of dual QCD is
\begin{equation}
\label{2eq18}
 E= M+\frac{\Omega_3r_0^4}{\kappa^2l^2},
\end{equation}
where $M$ is given in (\ref{2eq12}), and the entropy is still
given by (\ref{2eq13}). We can clearly see from (\ref{2eq18}) the
relation between the IR cutoff $r_0$ and the mass gap in the
holographic QCD..

As a result, we have seen that by introducing an IR cutoff, the
critical temperature for the corresponding Hawking-Page phase
transition increases, compared to the case without the cutoff. The
critical temperature is determined by the ratio $r_0/l$. In
addition, we stress here that the condition $r_0 <r_+$ in the
gravity side corresponds to the one that the mass gap is less than
temperature in the field theory side. Similar statement holds in
the following discussions.

\sect{Hawking-Page Phase Transition for Hyperbolic Black Holes with
an IR Cutoff}

In this section we will consider the so-called negative curvature
black holes whose horizon is a negative constant curvature surface.
In this case, the AdS black hole has the form
\begin{equation}
\label{3eq1}
 ds^2 =-V(r) dt^2 +V(r)^{-1} dr^2 +r^2 d\Sigma_3^2,
 \end{equation}
 where
 \begin{equation}
 \label{3eq2}
 V(r)= -1+\frac{r^2}{l^2}-\frac{r^2_s}{r^2},
 \end{equation}
 and $d\Sigma^2_3$ is the line element for the three-dimensional
 surface with curvature $-6$. With suitable identification, one can
 construct closed horizon surfaces with high genus. Such black holes
 are called topological black holes in some of the literature. In this case,
 the dual field theory resides on a hyperboloid described by $d\Sigma_3^2$.
  In this
 negative curvature case, there exist two remarkable features of
 the solution. One is that even when the mass parameter $r_s^2$
 vanishes, the metric (\ref{3eq1}) still has a black hole causal
 structure.  The metric function $V(r)$ is replaced
 by
\begin{equation}
\label{3eq3}
 V_b= -1 +\frac{r^2}{l^2}.
\end{equation}
 The black hole is called a ``massless'' black hole with a
 horizon at $r_+=l$. The other is that when $r_s^2$ is negative, there
 exists a ``negative mass" black hole with
 \begin{equation}
 \label{3eq4}
 V_b(r) =-1 +\frac{r^2}{l^2} +\frac{r^2_b}{r^2},
 \end{equation}
 provided $0< r_b^2 \le l^2/4$. When $r_b=l/2$, the black hole solution
 (\ref{3eq4}) is an extremal one with vanishing temperature.  For
 the black hole (\ref{3eq2}), the inverse Hawking temperature is
 \begin{equation}
 \label{3eq5}
 \beta = \frac{2\pi r_+}{2r_+^2/l^2-1},
 \end{equation}
 where $r_+$ is the black hole horizon satisfying $V(r_+)=0$. Once
 again, in order to regularize the Euclidean action, one has to
 choose a suitable reference background. One may consider the AdS space
 (\ref{3eq3}) as the reference background. But its shortcoming is that the period of the
 Euclidean time of the solution is fixed as $2\pi l$ and cannot be chosen as an arbitrary
 value,
 otherwise the reference background has a conical singularity.
 Another choice is the solution (\ref{3eq4}) with $r_b=l/2$.  In
 this case, the Euclidean time can be arbitrary since it describes
 an extremal black hole. Of course, when $r_b=0$, the solution
 (\ref{3eq4}) reduces to (\ref{3eq3}).  Therefore in what follows,
 we will consider the solution (\ref{3eq4}) with $r_b=l/2$ as the
 reference background. When $r_b=0$, it becomes the case
 (\ref{3eq3}). That is, in the following discussions,
 either $r_b=l/2$ or $r_b=0$.

 Considering the period $\beta_b$ of the Euclidean time for the reference background
 obeying
 \begin{equation}
 \beta_b\sqrt{V_b(R)}=\beta \sqrt{V(R)},
 \end{equation}
 at the boundary, and introducing an IR cutoff $r_0$ (here the meaning of $r_0$ is
 the same as the one in the previous section, the relation of $r_0$ to the mass
 gap in the dual field  theory can be obtained by calculating the mass from the
 Euclidean action below), we obtain
 the Euclidean action difference between the black hole and the
 background
 \begin{equation}
 \label{3eq7}
 {\cal I} = -\frac{\Sigma_3 \beta}{2\kappa^2 l^2} \left( 2r^4_{\rm
 max}-2r^4_0 -r_+^4+ r_+^2l^2 -r_b^2l^2\right),
 \end{equation}
 where $\Sigma_3$ is the volume of the closed horizon surface with unit
 radius. First let us consider the case without the cutoff. In
 that case, $r_{\rm max}=r_+$, and the action reduces to
 \begin{equation}
 \label{3eq8}
 {\cal I} =-\frac{\Sigma_3 \beta}{2\kappa^2 l^2} \left(r_+^4+ r_+^2l^2
 -r_b^2l^2\right).
 \end{equation}
Note that the minimal  horizon radius for the black hole
(\ref{3eq2})
 is $r_{\rm min}=l/\sqrt{2}$. If one chooses $r_b=l/2$ and
 (\ref{3eq4}) as the reference background, or $r_b=0$ and
 (\ref{3eq3}) as the reference background, then the action (\ref{3eq8})
 is always negative and the dual field theory is in the deconfinement phase.
  Therefore the usual Hawking-Page phase
 transition will not occur in this case.

 Next we consider the case with an IR cutoff $r_0$.  Here there are
 also two different cases.

 (i) $r_0>r_+$.  In this case, one has $r_{\rm max}=r_0$, and
 the action becomes
 \begin{equation}
 \label{3eq9}
{\cal I} = \frac{\Sigma_3 \beta}{2\kappa^2 l^2} \left(r_+^4-
r_+^2l^2 + r_b^2l^2\right).
 \end{equation}
 If $r_b=l/2$, we find that the action is always positive. Thus
 no phase transition occurs. If $r_b=0$, however, an interesting
 phenomenon appears. For those ``negative mass" black holes
 with horizon radius $l/\sqrt{2} \le r_+ <l$, the action is
 negative, while it is positive for black holes with horizon
 $r_+>l$.  The action changes its sign at $r_+=l$. This implies
 that in the low temperature phase $0<T<1/2\pi l$, the system is globally stable,
 while it becomes unstable as $T>1/2\pi l$; a Hawking-Page phase
 transition happens at $r_+=l$. This is an anti-intuitive result.
 This seemingly indicates that choosing the AdS black hole
 solution (\ref{3eq3}) is not suitable. Indeed, the result from
 the surface counterterm method indicates that one should choose
 $r_b=l/2$ as the reference background~\cite{EMP}.

(ii) $r_0<r_+$. In this case, we have $r_{\rm max}=r_+$, and the
action is given by
\begin{equation}
\label{3eq10} {\cal I} = -\frac{\Sigma_3 \beta}{2\kappa^2 l^2}
\left( r^4_+-2r^4_0 + r_+^2l^2 -r_b^2l^2\right).
 \end{equation}
 When $r^2_+ > r_c^2 $ with
 \begin{equation}
 \label{3eq11}
 r_c^2=
 \frac{l^2}{2}\left(-1+\sqrt{1+\frac{8r_0^4}{l^4}+\frac{4r_b^2}{l^2}}
 \right),
 \end{equation}
 the action is negative, while it is positive for $r_{\rm min}^2
 \le r_+^2 < r_c^2$.  The Hawking-Page phase transition happens at $r_+=r_c$.
 The critical temperature is
 \begin{equation}
 \label{3eq12}
 T_c= \frac{2r_c^2/l^2-1}{2\pi r_c}.
 \end{equation}
 Note that when $r_0=0$, one has $r_c<r_{\rm
 min}$, the action is always negative. In order to have $r_c
 >r_{\rm min}$, there exists therefore a gap for the IR cutoff
 $r_0$:
 \begin{equation}
 \label{3eq13}
 \frac{r_0^4}{l^4} > \frac{1}{8} \left( 3
 -\frac{4r_b^2}{l^2}\right)=\frac{1}{4}.
 \end{equation}
In the above calculation we have taken $r_b=l/2$. Therefore the IR
cutoff must be larger than the minimal horizon radius $r_{\rm
min}=l/\sqrt{2}$.

On the other hand, if we take $r_b=0$, the critical radius is
\begin{equation}
\label{3eq14} r_c^2=
 \frac{l^2}{2}\left(-1+\sqrt{1+\frac{8r_0^4}{l^4}}
 \right).
 \end{equation}
 The Hawking-Page phase transition happens when the temperature crosses the critical
 value (\ref{3eq12}). In this case, there is no gap for the IR cutoff
 $r_0$.

\sect{Hawking-Page Phase Transition for Charged AdS Black Holes with
an IR Cutoff}

In this section we consider the cases where the dual gravity
configurations are charged AdS black holes with different topology
horizons. In this case, the dual field theory is a Yang-Mills
theory with chemical potential at finite temperature.  We start
with the Einstein-Maxwell action with a negative cosmological
constant in five dimensions
\begin{equation}
\label{4eq1}
 {\cal S}= \frac{1}{16\pi G}\int d^5x\sqrt{-g} \left( R
 +\frac{12}{l^2}- F^2\right),
 \end{equation}
 where $F_{\mu\nu}$ is the Maxwell field  strength. This action can come
 from the spherical ($S^5$) reduction of type IIB supergravity~\cite{CEJM,CG,Cai}.
  The charged AdS black holes have the
 metric form
 \begin{equation}
 \label{4eq2}
 ds^2 =- V(r) dt^2 +V(r)^{-1}dr^2 + r^2 d\Omega_3^2,
 \end{equation}
 where
 \begin{equation}
 \label{4eq3}
 V(r) = k -\frac{m}{r^2} +\frac{q^2}{r^4}+\frac{r^2}{l^2},
 \end{equation}
 where $d\Omega_3^2$ is  the line element for a three-dimensional
 surface with constant curvature $6k$. Without loss of generality,
 one may take $k=1$, $0$ or $-1$. In addition, $m$ and $q$ are two
 integration constants, which are related to the mass and charge
 of the solution, respectively.

 The mass parameter $m$ can be expressed in terms of the
 horizon, $r_+$, as $m = r_+^2 (k +q^2/r_+^4+r_+^2/l^2)$.
 In order that the metric (\ref{4eq3}) describes a black hole with
 horizon $r_+$, the potential must have a positive radial derivative leading
 to the constraint on the horizon radius
 \begin{equation}
 \label{4eq4}
 2r_+^6+ kr_+^4l^2 -q^2l^2 \ge 0,
 \end{equation}
 otherwise, the solution describes a naked singularity. This
 condition can also be obtained from following Hawking
 temperature of the black hole (to keep the positiveness of
 the Hawking temperature).
 The Hawking temperature of the black hole is
 \begin{equation}
 \label{4eq5}
 T =\frac{1}{\beta} =\frac{1}{2\pi r_+}\left( k
 +\frac{2r_+^2}{l^2} - \frac{q^2}{r_+^4}\right).
\end{equation}
For the charged solution, the associated electric potential is
\begin{equation}
\label{4eq6}
 {\cal A} = \left( - \frac{q}{cr^2} +\Phi\right) dt,
 \end{equation}
 where $c=2/\sqrt{3}$, and $\Phi$ is a constant. We choose a gauge
 with $\Phi= q/cr_+^2$ so that the potential vanishes at the black
 hole horizon. Note that this provides us with a gauge invariant
 quantity because $\Phi$ describes a potential
 difference between at the horizon and at the infinity. Namely, if
one chooses another gauge,  $\Phi$ appearing in the following
equations represents the potential difference between at the
horizon and at the infinity. As a result, $\Phi$ is gauge
invariant.

 To discuss the thermodynamics and phase structure of dual QCD,
 one has to choose a suitable reference background.  For the
 charged solution (\ref{4eq3}), one suitable background is the
 solution (\ref{4eq2}) with
 \begin{equation}
 \label{4eq7}
 V_b(r) = k +\frac{r^2}{l^2} +\frac{r_b^2}{r^2}\delta_{k,-1},
 \end{equation}
 where $r_b=l/2$ or $0$, based on the discussion in the previous
 section.  The choice of the background implies that we are going
 to discuss the thermodynamics of the charged black holes in a
 grand canonical ensemble, where the temperature and electric
 (chemical) potential are fixed at the boundary. Note that although the
 solution (\ref{4eq7}) is the one without charge, we are
 discussing the thermodynamics in grand canonical ensemble, thus,
 the solution (\ref{4eq7}) with a constant potential $\Phi$ is
 still a solution of the action (\ref{4eq1}). Therefore the
 solution (\ref{4eq7}) with a constant potential is a good
 reference background, in order to analyze the thermodynamics of
 charged black holes in grand canonical ensemble~\cite{CEJM}.

Considering the solution (\ref{4eq7}) as the reference background,
we get the Euclidean action difference between the charged black
hole and the background
\begin{equation}
\label{4eq8}
 {\cal I}= \frac{\Omega_3 \beta}{16\pi Gl^2} \left( 2r_0^4
 -2r^4_{\rm max} +kr_+^2l^2 +\frac{q^2l^2}{r_+^2} +r_+^4
 -\frac{2q^2l^2}{r^2_{\rm max}}+ r^2_bl^2\delta_{k,-1}\right),
 \end{equation}
 where $\Omega_3$ is the volume of the three-dimensional closed
 surface $d\Omega_3^2$.

 We first discuss the case without the IR cutoff. In this case,
 $r_0=0$, and then $r_{\rm max}=r_+$. The action becomes
 \begin{eqnarray}
 \label{4eq9}
{\cal I} &=& -\frac{\Omega_3 \beta}{16\pi Gl^2} \left( r_+^4
-kr_+^2l^2 +\frac{q^2l^2}{r_+^2} - r^2_bl^2\delta_{k,-1}\right),
\nonumber \\
&=& -\frac{\Omega_3 \beta}{16\pi Gl^2} \left( r_+^4
-r_+^2l^2(k-c^2\Phi^2) - r^2_bl^2\delta_{k,-1}\right).
 \end{eqnarray}
 Obviously, when $k=0$ or $-1$, the action is always negative.
 Therefore the dual field theory is in the deconfinement phase and
 the Hawking-Page phase transition does not occur here. On the other
 hand, when $k=1$, the action changes its sign at
 \begin{equation}
 \label{4eq10}
 r_+^2 =r_c^2= l^2(1-c^2\Phi^2),
 \end{equation}
 provided $\Phi^2 <1/c^2$, which implies that the Hawking-Page
 phase transition happens at the critical temperature
 \begin{equation}
 \label{4eq11}
 T_{\rm HP}=\frac{3}{2\pi l}\sqrt{1 - c^2\Phi^2}.
 \end{equation}
 When $\Phi^2 >1/c^2$, the action is also always negative even if
 $k=1$. In that case, the dual field theory is in the deconfinement phase
 and the confinement/deconfinement phase transition will not
 occur.

 Next we consider the case with an IR cutoff $r_0$. And we will
 discuss separately the cases with $k=1$, $0$ and $-1$ below.

 \subsection{ Case $k=0$: Ricci flat black holes}

The dual field theory to this gravitational background is a
Yang-Mills theory at finite chemical potential and at finite
temperature living on a flat four dimensional spacetime. For the
gravitational computation in this case one has two subcases:
$r_0>r_+$ or
 $r _0<r_+$.

 (i) When $r_0>r_+$, one has $r_{\rm max}=r_0$. In this case, the
 action reduces to
 \begin{equation}
 \label{4eq12}
 {\cal I}= \frac{\Omega_3 \beta}{16\pi Gl^2} \left (r_+^4 +c^2\Phi^2 l^2r_+^2(1-2r_+^2/r_0^2)
 \right).
 \end{equation}
 The condition for the existence (\ref{4eq4}) of the horizon requires
 $r_+ >c\Phi l/\sqrt{2}$. We find that when $c\Phi l/\sqrt{2} <r_+
 <r_0$, the action (\ref{4eq12}) is always positive. As a result,
 no phase transition happens in this case.

 (ii) When $r_0<r_+$, we have $r_{\rm max}=r_+$, and the action
  is
  \begin{equation}
  \label{4eq13}
{\cal I}= \frac{\Omega_3 \beta}{16\pi Gl^2} \left( 2r_0^4
 -r^4_+ -c^2\Phi^2 l^2 r_+^2 \right).
 \end{equation}
 The action changes its sign from positive to negative  when the horizon radius
crosses
 \begin{equation}
 \label{4eq14}
 r_c^2 = \frac{-c^2\Phi^2 l^2+\sqrt{c^4\Phi^4l^4 +8r_0^4}}{2}.
 \end{equation}
 Note from (\ref{4eq4}) that there exists a minimal horizon radius
 $r_{\rm min}= c\Phi l/\sqrt{2}$.  we see from (\ref{4eq14}) that there is
 a gap for the IR cutoff. Namely the cutoff $r_0$ must satisfy
 \begin{equation}
 \label{in1}
 \frac{r_0^4}{l^4} > \frac{3}{8}c^4\Phi^4.
 \end{equation}
Clearly, this gap disappears when the charge is
absent~\cite{Herz}. The Hawking-Page phase transition
 temperature is
 \begin{equation}
 \label{4eq15}
 T_c= \frac{1}{2\pi r_c l^2}\left( -2 c^2\Phi^2 l^2
 +\sqrt{c^4\Phi^4 l^4 +8r_0^4}\right).
 \end{equation}
 Clearly, the phase transition
 occurs due to the introduction of the IR cutoff.

\subsection{Case $k=1$: spherical black hole}

(i) When $r_0>r_+$, one has $r_{\rm max}=r_0$ and the reduced
action is
\begin{equation}
\label{4eq16} {\cal I}=\frac{\Omega_3 \beta}{16\pi Gl^2} \left(
r_+^4 +r_+^2l^2 +c^2\Phi^2 l^2 r_+^2 (1-2r_+^2/r_0^2)\right).
 \end{equation}
 Considering the condition (\ref{4eq4}) for the existence of the black hole
 horizon, once again, we find that the action is always positive.
 Therefore, no phase transition will occur in this case.

 (ii) When $r_0<r_+$, we have $r_{\rm max}=r_+$, and the Euclidean
 action is
 \begin{equation}
 \label{4eq17}
{\cal I}= \frac{\Omega_3 \beta}{16\pi Gl^2} \left( 2r_0^4
 -r^4_+ + r_+^2l^2(1-c^2\Phi^2)\right).
 \end{equation}
From the action, we can see that the Hawking-Page phase transition
happens when the
 horizon radius crosses
 \begin{equation}
 \label{4eq18}
 r_c^2 =\frac{l^2(1-c^2\Phi^2) +\sqrt{(1-c^2\Phi^2)^2l^4
 +8r_0^4}}{2}.
 \end{equation}
 And the corresponding temperature at which the phase transition occurs is
 \begin{equation}
 \label{4eq19}
 T_c= \frac{1}{2\pi r_c}\left(
 2(1-c^2\Phi^2)+\sqrt{(1-c^2\Phi^2)l^4 +8r_0^4}\right).
 \end{equation}
 This temperature is higher than the one (\ref{4eq11}) for the
 case without the IR cutoff. Again, there is a gap for the IR
 cutoff because of the existence of the minimal horizon radius:
 the IR cutoff must be
 \begin{equation}
 \label{in2}
 \frac{r_0^4}{l^4} > \frac{3}{8} (1-c^2\Phi^2)^2.
 \end{equation}

\subsection{Case $k=-1$: hyperbolic black hole}

(i) When $r_0>r_+$, we have $r_{\rm max}=r_0$, and the action
reduces to
\begin{equation}
\label{4eq20} {\cal I}= \frac{\Omega_3 \beta}{16\pi Gl^2} \left(
r_+^4 -r_+^2l^2 +c^2\Phi^2 r_+^2 l^2 (1-2r_+^2/r_0^2)+
r^2_bl^2\right).
 \end{equation}
 According to the analysis in the previous section, we see that
 choosing the solution (\ref{3eq4}) with $r_b=0$ as the reference
 background is not reasonable. Therefore in this subsection we
 focus on the case with $r_b=l/2$. Once again, within the range
 $r_{\rm min}= l\sqrt{(1+c^2\Phi^2)/2} <r_+ <r_0$, the action is found to be always positive and
 no phase transition will occur, dual QCD is in the confinement
 phase.

 (ii) When $r_0<r_+$, one has $r_{\rm max}=r_+$. The action
 reduces to
 \begin{equation}
 \label{4eq21}
{\cal I}= \frac{\Omega_3 \beta}{16\pi Gl^2} \left( 2r_0^4
 -r^4_+-r_+^2l^2 -c^2\Phi^2 l^2 r_+^2+ r^2_bl^2\right).
 \end{equation}
 Clearly when $r_0=0$, no phase transition will occur. However,
 when $r_0 \ne 0$, the Hawking-Page phase transition happens when
 the horizon crosses $r_c$, satisfying
 \begin{equation}
 \label{4eq22}
 r_c^2=\frac{(1+c^2\Phi^2)l^2}{2} \left( -1 +\sqrt{1+\frac{8r_0^4
 +4r_b^2l^2}{(1+c^2\Phi^2)l^4}}\right).
 \end{equation}
 Since the critical radius must be larger than the minimal horizon
 radius $r_{\rm min}$, this leaves us with a gap for the IR cutoff
 \begin{equation}
 \label{4eq23}
 \frac{r_0^4}{l^4} > \frac{2+3c^2\Phi^2}{8}.
 \end{equation}
 Compared to the case (\ref{3eq13}) without charge, the IR cutoff
 gap increases.  Furthermore, the critical temperature of the
 phase transition is
 \begin{equation}
 \label{4eq24}
 T_c=\frac{1}{2\pi r_c}\left(-2 (1+c^2\Phi^2)l^2 +\sqrt{1+\frac{8r_0^4
 +4r_b^2l^2}{(1+c^2\Phi^2)^2l^4}} \right ).
 \end{equation}

\sect{Conclusions}

In this work we have studied the thermodynamics and Hawking-Page
phase transition of a hard wall AdS/QCD model in curved spaces by
introducing an IR cutoff, generalizing the work in \cite{Herz,BBBZ},
where the authors have discussed the phase transition between the
low temperature AdS space and the high temperature AdS black holes
in Poincare coordinates, which implies that the dual field theory
resides on a Ricci flat space. In our case, the dual field theory
lives in a curved space with positive or negative constant
curvature, dual to black hole configurations having spherical or
hyperbolic horizons.

In the case of the spherical AdS black holes, introducing the IR
cutoff leads the Hawking-Page phase transition temperature to
increase, compared to the case without the IR cutoff. For the case
of the hyperbolic black hole, the Hawking-Page phase transition will
not occur when one does not introduce an IR cutoff, while the
transition happens once the IR cutoff is introduced. However, there
is a gap for the IR cutoff in this case. Below that gap, the
Hawking-Page phase transition still does not occur.

For the charged AdS black holes with any horizon topology, like the
case without charge, the Hawking-Page phase transition becomes
possible due to the introduction of an IR cutoff. A remarkable
feature in this case is that for any horizon topology, a gap for the
IR cutoff always exists due to the existence of a minimal black hole
horizon.

\section*{Acknowledgments}
RGC thanks C. Herzog for helpful email correspondence and Y.Q. Chen,
C. Liu, J.P. Ma  and F. Wu for useful discussions.  The work of
R.G.C. was supported in part by a grant from Chinese Academy of
Sciences, and by the NSFC under grants No. 10325525 and No.
90403029. The work of J.S was funded by the NSFC under grants No.
10475105, 10491306 and PHY99-07949.

\end{document}